**Reconstructing data-driven governing equations for cell phenotypic transitions: integration of data science and systems biology**

Jianhua Xing[1,2,3,*]

[1] Department of Computational and Systems Biology, University of Pittsburgh, Pittsburgh, PA 15232, USA.

[2] Department of Physics and Astronomy, University of Pittsburgh, Pittsburgh, PA 15232, USA.

[3] UPMC-Hillman Cancer Center, University of Pittsburgh, Pittsburgh, PA, USA.

* To whom correspondence should be addressed. Email: *xing1@pitt.edu*

**Abstract**

Cells with the same genome can exist in different phenotypes. and can change between distinct phenotypes when subject to specific stimuli and microenvironments. Some examples include cell differentiation during development, reprogramming for induced pluripotent stem cells and transdifferentiation, cancer metastasis and fibrosis development. The regulation and dynamics of cell phenotypic conversion is a fundamental problem in biology, and has a long history of being studied within the formalism of dynamical systems. A main challenge for mechanism-driven modeling studies is acquiring sufficient amount of quantitative information for constraining model parameters. Advances in quantitative approaches, especially high throughput single-cell techniques, have accelerated the emergence of a new direction for reconstructing the governing dynamical equations of a cellular system from quantitative single-cell data, beyond the dominant statistical approaches. Here I review a selected number of recent studies using live- and fixed-cell data and provide my perspective on future development.

**Conflict of Interest declaration:** The authors declare that they have NO affiliations with or involvement in any organization or entity with any financial interest in the subject matter or materials discussed in this manuscript.

**Introduction**

Cells are fundamental units of living organisms. Shared with the same set of genome, a cell can exist in different cell phenotypes with distinct physical and biological features. For example, bacteria like anthrax can stay in either a dormant state for years, or an active state. Cells of higher organisms such as mammalian cells can exist in more phenotypes such as embryonic stem cells, neurons, fibroblasts, etc. Furthermore, cells can convert between different phenotypes either spontaneously or being induced. Examples of the former include embryo development, wound healing, and pathological processes including cancer metastasis and fibrosis. One known example of the latter is reprogramming terminally differentiated cells into induced pluripotent stem cells or other cell types, which opens a new direction for regenerative medicines (1-5). Therefore, quantitative and mechanistic understanding of a phenotypic conversion process is of fundamental biomedical significance for effective control of the phenotype of a cell.

A cell is a complex system composed of many molecular species that interact with each other to form a regulatory network. The network regulates various cellular functions such as uptaking extracellular chemical molecules as food, receive chemical and mechanical signals, and making cell fate decisions. Extensive efforts have been made to decipher how a regulatory network, formed through a collection of coupled biochemical reactions, controls cellular dynamics, especially cell phenotypes (6-9). Mathematically, a cell is a dynamical system and the temporal evolution of the cell state can be described by a set of dynamical equations. There are extensive mathematical modeling and experimental studies on the multistability of biological systems (10-20). Consequently, different cell phenotypes correspond to different stable attractors of a nonlinear dynamical system, and a phenotypic transition is the transition from an (at least originally) stable attractor to another one. Transition dynamics between two (equilibrium or nonequilibrium) stable attractors is a classical theoretical question in applied mathematics, physics and chemistry, and continues to be an active research area even after more than a century of continuous studies (21).

A central step in mathematical modeling studies of cell state transitions is to construct the governing equations of the cellular dynamics, then one can perform either theoretical analyses such as bifurcation and sensitivity analyses, or numerical simulations. For example, in a typical systems biology study, one formulates the governing equations based on principles of physics and chemistry, and biological knowledge of the system, then uses quantitative data to constrain the model parameters (22). This framework is flexible on incorporating qualitative and quantitative information about different layers of regulation, e.g., epigenetics, transcription, translation, post-translation, etc., which may be obtained from various experimental approaches. A main limitation is that most of this type of model studies are restricted to small networks or a few degrees of freedom (6, 7). However, in reality simple networks are embedded in a much larger densely-connected network, and it is highly nontrivial to reconstruct the dynamics of such a network. Although efforts have been made to perform whole-cell simulations of bacteria (23, 24), it remains a grand challenge to reconstruct similar models for more complex cells from experimental data.

Recent years advances of single cell techniques, both for fixed and live cells, facilitate generation of a large body of single cell data (25-29). Together with advances of machine-learning-based computational approach development, an emergent research direction is to extract dynamical and mechanistic information from "big data" for cellular processes. Here my goal is to provide a brief overview of some recent developments, and my personal perspective.

**Formulation of cell phenotypic transitions in the context of dynamical systems theory**

Let's first formulate the problem of dynamics reconstruction from data in a general form. Assume that a state vector **z** completely specifies the internal cellular state at a given time. Mathematically completeness of the state representation means here that one can write down a set of memory-less Langevin equations that govern the dynamics of individual cells (30),

$$d\mathbf{z}/dt = \mathbf{A}(\mathbf{z}, \lambda) + \zeta(\mathbf{z}, t), \qquad (1)$$

where $d\mathbf{z}/dt$ is the instant velocity of a cell moving in the state space, **A** is a governing vector field containing information on how system components interact and act on each other at a given time, λ refers to controlling parameters and extracellular environmental factors that in general can be time-varying, and **ζ** are white Gaussian noises with zero mean to account for fast dynamics of degrees of freedom not explicitly described. Notice Eqn. 1 alone describes how the cellular state of one cell evolves, and a separate description is needed for the change of the cell ensemble due to cell birth and death. Then, there are two basic questions for reconstructing the dynamics of a cellular system excluding the additional birth-death events: what is the appropriate choice of the dynamical variable **z**, and how to get the function **A** and **ζ**.

For a cellular system, in principle one choice of **z** is the compartmentalized concentrations (or amount) of all intracellular species. In practice, however, one only knows information of a subset of the species. For example, with single cell RNA-seq data one can represent a cell state as a vector in the genome-wide transcriptomic expression space. Alternatively, collective cell features, such as morphological and cell texture features, have been widely used for phenotyping cells (31-37). Accumulating studies demonstrate that cell morphological features faithfully reflect the underlying cell expression states. Barkal et al. showed that cell morphological features define the corresponding regulatory network (38). Yin et al. showed with multiple cell lines that gene regulation leads to switch-like transitions between discrete cell shapes (35). Gordonov et al. developed an experimental-computational framework to characterize cell morphological changes in response to drug treatments (31). Wu et al. demonstrated that cells with different morphology have distinct expression profiles and metastasis potential (39). While in most studies these composite cell features are treated as statistical quantities for clustering cells, recent studies have treated them as dynamical variables to specify cell states (31, 40-44), analogous to using collective modes in condensed matter physics. In all cases one quantifies measurements, either sequencing reads or images, into a vector representing cell state in a multi-dimensional state space. In the following discussions, we will use **z** for a generic cell state, **x** in the expression space, and **y** in the composite cell feature space.

For a gene regulatory network, **A(z)** provides a quantitative description about gene regulation. For cell feature representation, **A(z)** gives coupling between different cell features, which reflects the underlying interactions of involved cellular species. Theoretically, assume that one can record an infinite long ergodic stochastic trajectory of **z**(t), then $A(z) = \lim_{\Delta t \to 0} \frac{1}{\Delta t} \{\lim_{t \to \infty} \frac{1}{t} \int_0^t ds(z'(s+\delta t) - z'(s))\delta(z'(s) - z)\}$, and $\zeta(z,t)dt = \sqrt{2D(z)}dW(t)$, where $W(t)$ is a multidimensional Wiener process, and $2D_{ij}(z) = \lim_{\Delta t \to 0} \frac{1}{\Delta t} \{\lim_{t \to \infty} \frac{1}{t} \int_0^t ds(z_i'(s+\delta t) - z_i'(s))(z_j'(s+\delta t) - z_j'(s))\delta(z'(s) - z)\}$. However, cellular dynamics has a broad distribution of time scales, and it is generally impractical to record long (i.e., multiple-cell-cycle) single cell trajectories. So, it is more feasible to replace time averages with ensemble averages. To find a numerical approach of reconstructing **A(z)**, let us examine the Fokker-Planck equation corresponding to the above Langevin equations (with the assumption that λ is time-invariant for simplicity),

$$\frac{\partial n(z,t)}{\partial t} = \nabla \cdot [-J(z)n(z,t) + D \cdot \nabla n(z,t)] + \int dz' B(z,z')n(z',t),$$

(2)

where $n(z,t)$ is the number density. The last nonlocal term accounts for discrete changes due to cell division, and its diagonal terms also include contribution from cell death. One expects that the function **B(z,z')** is peaked in restricted regions populated with cells right before and after division. The term **J** is related to the vector field under the Stratonovich interpretation of the stochastic differential equations (45, 46) as,

$$J(z) \equiv \frac{\int dz' n(z',t)\frac{dz'}{dt}\delta(z'-z)}{\int dz' n(z',t)\delta(z'-z)} = A(z).$$

(3)

The above relation of **J** immediately suggests one algorithm to reconstruct **A(z)** from measured (**z**, d**z**/dt). A limitation of this algorithm is that it requires information of $dz/dt$, which can be challenging for fixed cell data (but see below). In practice one typically replaces the Dirac's δ function with a finite volume bin and performed the average over $dz/dt$. Notice that $n(z,t)$ is expected to be a slowly varying function over the scale of the bin size, so the evaluation of **J(z)** is insensitive to the sampling of $n(z,t)$. This is an attractive feature as compared to the algorithm discussed below.

Another possible algorithm is to regress the equations from a series of measurements of $n(z,t)$ at different time points. This algorithm does not require information of $dz/dt$, but is sensitive to the sampling accuracy of the multi-dimensional density function $n(z,t)$, which is challenging. A further technical challenge is that in most single cell experiments one typically measures the cell probability density $\rho(z,t) = n(z,t)/N(t)$ with N being the total cell number instead of the number (or concentration) density. Then, when N(t) changes with time, unless one also measures $N_t$ to obtain $n(z,t)$, the partial differential equation of $\rho(z,t)$ contains nonlinear terms. For the convenience of seeing how the nonlinear terms arise, and for practical applications, let us examine a discretized version of Equation 2. We divide the coordinate space into sufficiently small regions, and one can convert the partial differential equations into, or formulate directly from the continuity condition that also applies to systems with dynamics more general than that described in Eqn. 1 and 2 such as discrete state dynamics, a set of master equations,

$$\frac{dn_i}{dt} = \underbrace{\sum_{<i,j>}\left(-k_{ij}n_i + k_{ji}n_j\right)}_{\text{continuous cell growth and cellular responses}} + \underbrace{\left(-\alpha_i n_i + \sum_{j\neq i}\beta_{ji}n_j\right)}_{\text{cell division}} - \underbrace{d_i n_i}_{\text{cell death}}, \quad (4)$$

or (see also (47)),

$$\frac{dp_i}{dt} = \sum_{<i,j>}\left(-k_{ij}p_i + k_{ji}p_j\right) + \left(-\alpha_i p_i + \sum_{j\neq i}\beta_{ji}p_j\right) - d_i p_i - p_i \sum_j (a_j - d_j)p_j, \quad (5)$$

where $n_i$ and $p_i$ are the number of cells and the fraction of cells in region $i$ of the state space, respectively. The terms in the first parenthesis are the flux exchange between two neighboring regions due to cell growth or cellular responses to stimuli. The terms in the second parenthesis are due to jumps between often non-neighboring regions in the state space due to cell division, and a relation $2\alpha_i = \sum_{j\neq i}\beta_{ij}$ holds to account for the fact that one mother cell divides into two daughter cells. The third term accounts for change of $n_i$ from cell death. Notice that the last term of Eqn. 5 is nonlinear in $p$.

**Extraction of dynamical information from snapshot data**

Given the destructive nature of many single cell techniques, continuous attempts have been made on extracting dynamical information from snapshot data. In this section I will discuss a few selected recent developments.

A class of methods use the ergodic principles (48). The approaches are based on some basic assumptions: 1) one measures the stationary distribution of the underlying system; 2) the average of the dynamics of one cell over time is equivalent to the average over a population of cells at one time point. One example is the ergodic rate analysis (ERA) method of Kafri et al. (49). They applied ERA to analyze cell size regulation of mammalian cells under the exponential growth condition. Under this condition the total cell number of the population grows exponentially, $N_t = N_0 \exp(\alpha t)$, and they deduced the growth rate from measuring $N_t$ at different time points followed by exponential fitting. Then they measured the stationary probability density distribution the two-dimensional state space defined by DNA content and the activity of the anaphase-promoting complex (APC) in its active form using fluorescent staining. To simplify the ERA analysis, they further projected the two-dimensional stationary distribution to an average one-dimensional cell cycle axis describing cell cycle progress through $G_1 \rightarrow G_1/S \rightarrow G_2$ then back to $G_1$ through cell division. Divide the cell cycle axis into discrete regions, with $p_i$ the measured percentage of cells in region $i$, and $P_i$ the cumulative cell fraction, both can be obtained from the fixed cell data. Then Eqn. 4 reduces to a simple Markov model shown in Fig. 1A, and the flux balance condition leads to $\frac{dN_i}{dt} = \alpha P_i N_t = Flux_0 - Flux_i = 2\alpha N_t - \omega_i p_i N_t$, so $\omega_i = \alpha \frac{2-P_i}{p_i}$. The factor 2 in $Flux_0$ comes from that one mother cell divides into two daughter cells.

From such ERA analyses Kafri et al. identified that cell size feeds back onto cell cycle regulation.

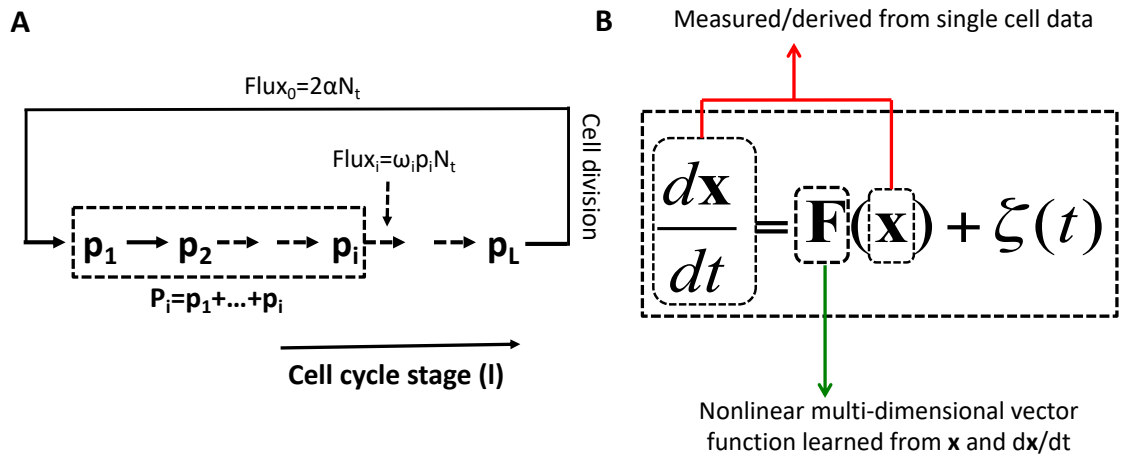

*Figure 1 Reconstructing quantitative models from single cell snapshot data. (A) A Markovian model of cell cycle progression based on ergodic rate analyses. (B) Reconstruction of genome-wide vector fields from scRNA-seq data (50). The vector **x** represents the cell expression state, which refers **s** in the splicing-based velocity, and **(u + s)** in the more accurate metabolic labeling-based velocity. The term ζ refers to residue terms treated as random noises. The function **F** corresponds to **A** in Eqn. 1.*

Weinreb et al. applied similar idea to analyze scRNA-seq data and termed their method population balance analysis (PBA) (51). Here again one assumes that the system is under steady state. Since the dimensionality of the state space is typically high, it is impractical to use regular grids. Instead they used spectral graph theory to represent the multi-dimensional Fokker-Planck equations on the irregular grids formed by the data points. Furthermore, to simplify the determination of the function form of **A**, they assume that it can be written as the gradient of a scalar potential, $\mathbf{A} = \nabla \phi$. Rigorously speaking this assumption is not complete for describing the dynamics of a system without detailed balance such as a cellular system. Weinreb et al. performed systematic analyses on the limitations of extracting dynamical information from snapshot data under this framework. Furthermore, Klein and coworkers showed combining snapshot data and lineage tracing provides a more reliable strategy for reconstructing the genome-wide transcriptional trajectories (52, 53).

With time series measurements using fixed cells, several studies reconstructed cellular dynamics without the steady-state requirement. Yeo et al. introduced a generative model called PRESCIENT (Potential eneRgy undErlying Single Cell gradIENTs) to learn differentiation landscape from time-series scRNA-seq data (54). They also assume that $\mathbf{A} = \nabla \phi$, and use deep learning to determine the function form of ϕ by fitting the calculated **n(z**,t) to the measured ones. With ϕ and the governing dynamical equations, one has a generative model and can perform in silico perturbation studies. PRESCIENT incorporates cell growth rates derived from experimental lineage tracing or approximated from growth hallmark gene expression, and such incorporation significantly improves the model prediction accuracy. They demonstrated that PRESCIENT trajectories perform better on the ability to recover experimental lineage tracing from hematopoiesis, and achieve a performance on cell fate prediction, i.e., predicting the terminal fate of lineage traced progenitors, superior to a few other methods including dynamo discussed below. With the framework one can also perform single and multiple gene in silico perturbations, and the predictions agree nicely with the well-characterized perturbations along the axis of endocrine development.

Waddington-OT is another strategy using time series scRNA-seq data that has received lots attention (55). The method formulates the temporal evolution from a measured state space number density to that of a subsequent time point as an optimal transport problem to determine a transition matrix corresponding to that in Eqn. 4, with a generalization that the transitions may take place between non-neighboring regions in the cell state space since the time difference between the two densities is finite. Thus the transition matrix here can be related to a Green's function. Notice here differing from a conventional optimal transport formulation, cell densities instead of normalized distributions are used to account for cell birth and death. A central part of Waddington-OT is to choose the cost function of "transporting" cells from one state to another one. While in the original study a simple square function of the Euclid distance between two cell states is used, it is desirable to relate the transport cost function to explicit biological and physical constraints of the cellular dynamics, e.g., using a distance weighted by gene-specific dynamics.

In another set of applications, one clusters cell based on expression or other cell features obtained through sequencing, immunostaining, imaging, etc, and measures how the numbers (i.e., $n_i$ in Eqn. 4) or fractions (i.e., $p_i$, in Eqn. 5) of individual subpopulations change over time. Then one fits the data with a model of transitions between these discrete states, with the Markovian dynamics often (but not always) assumed. Some examples include spontaneous transitions among cancer cell subpopulations (56, 57), and induced cell phenotypic transitions through intermediate subphenotypes (44, 58-60).

While all the above approaches are based on the measured distributions, Qiu et al. developed a framework to learn a set of high-dimensional analytical vector field functions from sparse, noisy, and discrete scRNA-seq datasets (50). To explain the basic idea, let's start with the original RNA velocity estimation approach, which estimates instant RNA transcriptional rates from the spliced and unspliced RNA reads in an scRNA-seq experiment (61). The conventional RNA velocity method from the original paper is based on a generic transcription model for a gene $i$: first, transcription (with a rate constant $α_i$) leads to unspliced mRNA (with $u_i$ specifying the copy number); the unspliced mRNAs are converted to spliced mRNAs ($s_i$) through a splicing process that is approximated as a one-step rate process with a rate constant $β_i$,; the spliced mRNAs are eventually degraded with a degradation constant $γ_i$. The overall mRNA turnover dynamics then can be described by a set of ordinary differential equations (ODEs) as follows:

$$\frac{du_i}{dt} = α_i - β_i u_i, \frac{ds_i}{dt} = β_i u_i - γ_i s_i.$$

Notice that for each gene one can read out $u_i$ and $s_i$ directly from scRNA-seq data. By taking $β_i$ = 1 (so all other rate constants are values relative to splicing instead of absolute values), and finding ways to estimate $γ_i$, La Manno et al. showed that one can estimate RNA velocity for this gene, d$s_i$/d$t$ = $u_i$ − $γ_i s_i$. Notice that the velocity depends on both $u_i$ and $s_i$. A positive velocity means that this gene is under upregulation, while a negative value means downregulation. Therefore, for a cell at a given expression state (represented by **s**, a vector with each component representing the copy number of spliced RNA of a gene), another velocity vector **v** = d**s**/d$t$ gives the direction and speed this cell moves in the expression state space. The splicing-based RNA velocity estimation has several inherent limitation on its accuracy. Recently Qiu et al. generalized RNA velocity estimation to scRNA-seq data with RNA metabolic labeling (50). RNA metabolic labeling treats live cells with certain chemicals to incorporate modified ribonucleosides, such as 5-Bromouridine (BrU) or 4-thiouridine(4sU), into nascent RNA transcripts. At the sequencing stage, the modified ribonucleosides induce ribonucleoside exchange with certain probabilities that can be captured through sequencing. From the amount of nascent and old (unlabeled) transcripts one can infer genome-wide RNA turnover dynamics (62-66), then one can estimate RNA velocities that are both absolute and have improved accuracy over the splicing-based method.

A subtle but important conceptual advance of Qiu et al. is to realize that the rate "constants" $α_i$, $β_i$, and $γ_i$ are really functions of concentrations of molecular species in the cell. For example, $α_i$ is a function of the concentrations of transcription factor proteins that regulate gene $i$. Therefore, let's make the following assumptions: 1) the state of a cell can be specified by the current transcriptomic state, say the spliced RNA vector **s**; 2) the transcriptional regulation is a function of **s**. The first assumption is made when we only have transcriptomic data, and one can relax it when multi-omics data become available. The second assumption is due to the nature of snapshot data, and relaxing it would generally require true live-cell time series data discussed below. Then the unspliced mRNA vector **u** = **u**(**s**) is also a function of **s**, and one reaches that, $\frac{ds_i}{dt} = β_i u_i(s) - γ_i s_i = F_i(s)$, where the function **F** forms a to-be-determined genome-wide, generally nonlinear vector field in the expression state space defined by **s**, and it contains quantitative gene-gene regulation formation.

Qiu et al. developed a machine learning procedure, called dynamo, to reconstruct the **F** function from scRNA-seq data (Fig. 1B) (50). Given the scarcity of the data and the high dimensionality of the system, it is numerically challenging to apply Eqn. 3 directly. Instead, from measured **x** and d**x**/dt, one learns the function form of **F** that maps **x** to d**x**/dt as a regularized optimization problem in an infinite dimensional function space, technically called a Reproducible Kernel Hilbert Space (RKHS). That is, one assumes that **F** can be expressed as a linear combination of $M$ basis functions, $F(x) = \sum_{m=1}^{M} C_m \Gamma_m(x)$, then the learning problem reduces to determine the coefficient vectors $C_m$. A prominent feature of the procedure is that it learns an analytical and continuous function form of the vector field so one can obtain **F** for regions unexplored by the sampling data. Therefore, dynamo gives a generative model that allows further in silico studies, which is similar to PRESCIENT but with significantly more information, i.e., RNA velocities besides gene expression, as input, thus abolishing the need of assuming that the dynamics is governed by the gradient of a scalar function. The function **F** contains quantitative information on gene regulation. Qiu et al. have developed a number of analysis tools, including efficient differential geometry analyses exploiting the analytical function form of the transcriptomic vector field, and using least-action path (LAP) analysis to predict optimal transition paths between two distinct phenotypes and key regulatory factors for the transition processes. Qiu et al. applied to a dataset on spontaneous differentiation of primary human CD34+ hematopoietic stem cells (HSPCs), and accurately revealed genome-wide quantitative regulation relations between genes, such as the mutual antagonism between two master regulators GATA1 and SPI1/PU.1, from a single set of high-quality scRNA-seq data. Traditionally considerable efforts are needed to acquire such quantitative information even for a pair of genes (67). Further in silico LAP analyses recovered key reprogramming factors that have been experimentally reported to lead reprogramming between different cell types.

All these methods are designed to extract dynamical information from static data, and technically they differ on whether one focuses on individual cells or cell populations, similar to particle-based and field-based descriptions of a physical system. With infinite amount of information about the system, mathematically both population and individual-based approaches converge. In practice, only limited experimental information is available at various levels, so these methods use different strategies to constrain parameters. With population information at some finite temporal and state resolution, population-based methods such as PRESICENT and Waddington-OT set some *a priori* constraints such as assuming gradient dynamics. With gene-specific transcriptional dynamics estimated based on a mathematical framework of rate equations, theoretically dynamo can provide significantly more details on how genes are regulated, as demonstrated by the example on GATA1 and SPI1/PU.1. In practice, the accuracy of the reconstructed vector field depends on the quality of the data, and future efforts should be especially put on more accurate estimation of gene expression dynamics.

These approaches are not exclusive to each other. Specifically, the framework of dynamo is general and flexible for incorporating additional information. For example, if time series scRNA-seq data is available, one may follow the dynamo procedure illustrated in Fig. 2B to learn the form of **J** in Eqn. 2 that describes the deterministic part of intracellular network dynamics of one cell, adopt the procedure of PRESICIENT to estimate the cell birth/death rates, i.e., the **B** matrix, then estimate the diffusion matrix **D** by minimizing the error between the distributions at various time points calculated from Eqn. 2 and the experimental ones. With the single cell vector fields, one can also perform multi-scale agent-based modeling to simulate individual cells, and it is straightforward to include interactions between cells and between a cell and extracellular environmental factors.

Given the capacities of dynamo, the success of methods like PRESCIENT and WADDINGTON-OT does raise a profound question on what level of details is or is not needed to describe certain aspects of a cellular process. A similar situation has been encountered in studying macromolecule dynamics. A class of coarse-grained models, elastic network models, which represent a protein or its complexes/assemblies as a collection of nodes connected by springs, with spring constants either being universal (68, 69) or having a simple distance dependence (70), perform remarkably well on revealing structure-function relations. Notably, these models neglect physicochemical differences between building blocks such as the identities of amino acids. Yet, they provide an accurate description of the cooperative (often functional) dynamics of the overall structure (71). This suggests that the overall contact topology and associated entropic effects play a dominant role in defining the accessible relaxational motions, which, in turn, are recruited for achieving function. Therefore, there is no need to include detailed force fields or amino acid specificity if the goal is to characterize the most collective motions of the overall architecture. Such features become important in local events/interactions, however. It remains to see whether one can reach similar conclusions for cellular dynamics.

### Extraction of dynamical information from live-cell time series data

Despite the success of extracting dynamical information from snapshot data, the long-term dynamical information can only be inferred indirectly with some assumptions whose validity has not been systematically evaluated. For true long-term dynamics one needs to perform live-cell imaging over the same cells over time, which has its unique technical challenges. Here we specifically focus on cell phenotypic transition dynamics of eukaryotic cells, which typically takes days or weeks. Compared to fixed cell studies, there are more constraints on choosing observables since the number of measurable degrees of freedom is much more restricted. Monitoring these observables should be compatible for live-cell imaging with minimal cell toxicity and perturbation to the cellular dynamics, and these observables should be sufficient to reflect the course of the transition process. Suppose that a gene is activated and reaches a plateau at the early stage of the transition, monitoring it alone is less informative on the later stage of the transition. However, cell- and photo- toxicity as well as the number of available fluorescent channels limit the number of molecular species one can monitor for characterizing cell phenotypic transitions using fluorescent-based labeling. For example, studies on the epithelial-to-mesenchymal transition (EMT) lead to a consensus that it is insufficient to use a small number of markers to reflect the progression of the process (72). Collective cell features measured through label-free imaging, either alone or in combination with fluorescent imaging, arise as alternative quantities for live-cell imaging.

While live-cell imaging studies on cell phenotypic transitions are still at an early stage, recent years witness an increasing number of such studies. Here I will only discuss a few of them. Buggenthin et al. showed that a combination of cell morphological features and movement accurately predicts lineage choices of mouse primary hematopoietic progenitors three generations before indication by conventional molecular markers (73). In a series of studies (42, 43), Marshall and coworkers examined multiple types of cells in cell motility and morphology state space. Figure 2A shows the measured transition vectors of mouse embryonic fibroblasts (MEFs) in the state space defined by the two leading two principal components (PCs) of a 205-dimensional cell size, shape, and texture feature space. Flux analyses on the transitions reveal that the forward and backward transitions between two states (see Eqn. 4) satisfy detailed balance (Fig. 2B), which is not expected for a cellular system that is out of thermodynamic equilibrium. With detailed balance one can define a scalar potential to represent the system dynamics, which reveals a multi-dimensional attractor in the state space (Fig. 2C).

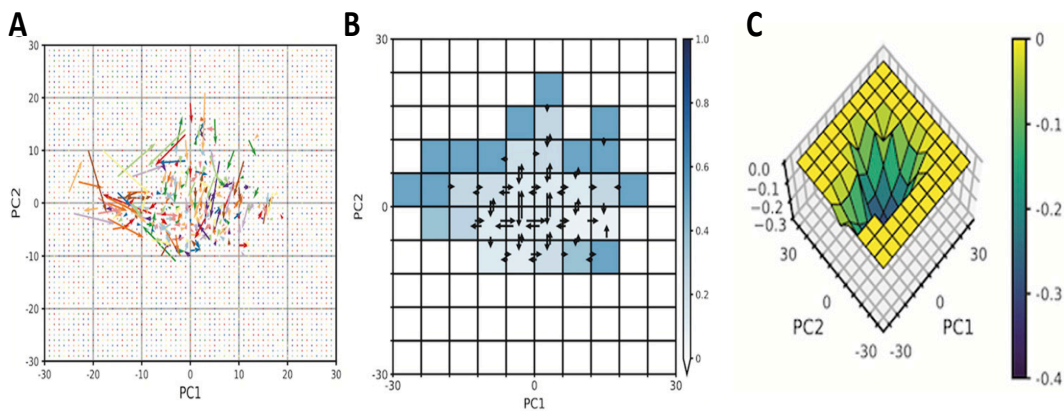

*Figure 2 Cytomorphological state space dynamics of mouse embryonic fibroblasts (MEFs). (A) Transition vectors in the leading PC space. The vectors are defined as the vector between the measured states of individual cells with four-hour separation. (B) Flux analyses of the transition vectors reveal a transition matrix with unexpected detailed balance between forward and backward states. (C) Quasi-potential defined from the transition matrix. Reproduced from (43) with permission.*

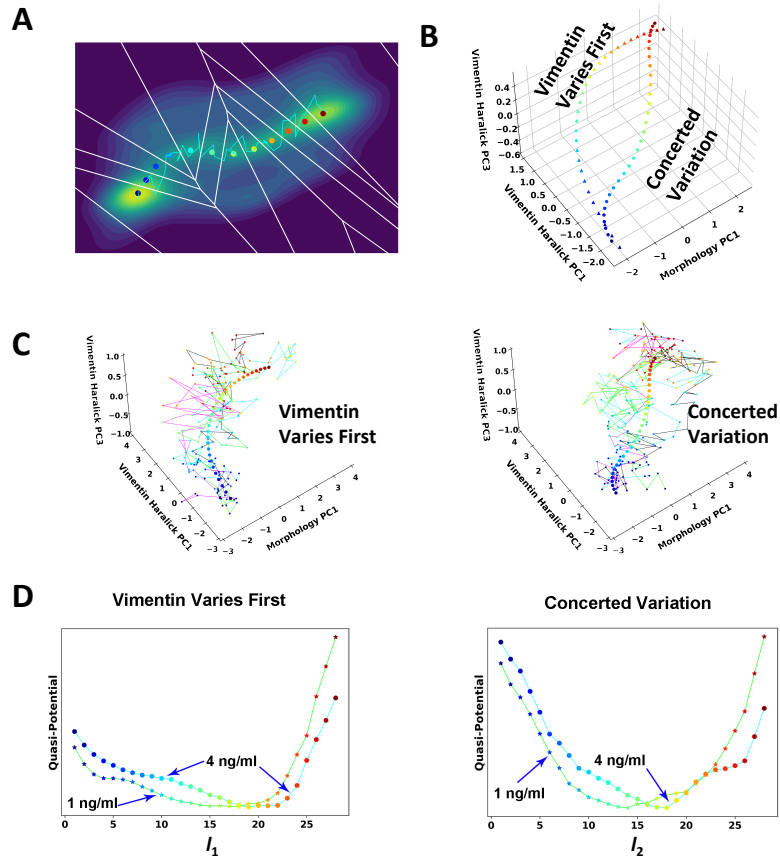

*Figure 3 Determination of reaction coordinates from single cell trajectories using a revised finite temperature string method. (A) Division of cell state space with Voronoi cells. A, I, and B refer to the initial, intermediate, and final regions of the transition, respectively. (B) Reconstructed parallel EMT transition paths of A549 cells (treated with 4 ng/ml TGF-β) shown in the leading 3-D state space. (C) Individual reaction coordinates overlapped with typical recorded single cell trajectories. (D) Reconstructed quasi-potentials along reaction coordinates from single cell trajectory data of A549 cells (treated with 1 and 4 ng/ml TGF-β). Reproduced from (40).*

Wang and co-workers developed an integrated live-cell imaging and image analysis platform for studying cell phenotypic transitions (40, 41, 74). Through representing cells in a composite cell feature space, they obtained single cell trajectories, then applied rate theories to analyze the transition dynamics. Rate theories are an old branch of physics studying how a system transits between two distinct regions (i.e., regions A and B in the schematic diagram of Fig. 3A) in a state space (21). In rate theories, a trajectory is called a reactive one if it leaves region A and ends at region B before returning to A, and all of them form an ensemble of reactive trajectories. There are an infinite number of reactive trajectories with varying probabilities that connect A and B, and they typically concentrate within a "reaction tube" (75, 76). Furthermore, reaction coordinate (RC, denoted as *l* in Fig. 3) is a key concept in rate theories, and it refers a one-dimensional manifold that connects A and B. Here I used the same notation as in Fig. 1 since the cell cycle stage there can be viewed as a RC. One way of defining a RC is the centroid of the reaction tube, and one may reconstruct such a RC from the reactive trajectory ensemble using a finite temperature string method (40, 75). Figure 4B shows two RCs corresponding to two parallel paths for the EMT process of A549 cells treated with TGF-β. Recorded single cell reactive trajectories fluctuate around these RCs and form the reaction tubes (Fig. 3C). With both the cell states and instant velocities from live-cell imaging trajectories, one can apply Eqn. 3 to reconstruct the vector field **A**. For simplicity and better numerical convergence, Wang et al. considered the equation projected along each 1-D RC $l_\alpha$, which then reduces to the form, $\frac{dl_\alpha}{dt} = -\frac{d\phi_\alpha}{dl_\alpha} + \zeta_\alpha(l_\alpha, t)$, where α = 1, 2 for the two parallel RCs. Here since the equation is one-dimensional, one can always define a scalar quasi-potential $\phi_\alpha(l_\alpha)$ for each RC. Numerically, the discretized RCs divide the multi-dimensional state space into Voronoi cells, so Eqn. 3 becomes, $\frac{d\phi_{\alpha i}}{dl_\alpha} = -\langle\frac{dl_\alpha}{dt}\rangle_{i-th\ Veronoi\ cell}$, with the average over all measured (reactive and non-reactive) trajectories that fall into the specific Voronoi cell. Then the quasi-potential relative to a reference point ($l_{\alpha 0}$) can be obtained through integration, $\phi_{\alpha i} - \phi_{\alpha 0} = \int_{l_{\alpha 0}}^{l_{\alpha i}} \frac{d\phi_\alpha}{dl_\alpha} dl_\alpha \propto \sum_{j=1}^{i} \frac{d\phi_{\alpha i}}{dl_\alpha}$. The position-dependent diffusion "constant" is given by, $D_{\alpha i} = Variance\left(\frac{dl_\alpha}{dt}|_{\alpha i}\right)$. The reconstructed quasi-potentials (Fig. 3D) reveal that TGF-β destabilizes the original epithelial attractor and the EMT process proceeds as relaxation to a new attractor. The process is analogous to a molecular system being excited from an electronic ground state potential and relaxing along the potential of an excited state (77). The remnant of the epithelial attractor appears as a plateau in the quasi-potential of one path (Fig. 3D left), and the attractor is more apparent with a lower TGF-β concentration. The Markovian assumption was justified with the Chapman-Kolmogorov test. The reconstructed Fokker-Planck equation predicts a stationary distribution that agrees well with experimental results, which is impressive given that the dynamical equation was reconstructed from non-stationary trajectories. With more data one may also follow the procedure of Qiu et al. (50) to obtain an analytical form of the multi-dimensional vector field from live-cell imaging data.

## Challenges and perspectives

Cell biology studies are in an exciting "big data" era. While statistics-oriented approaches are still dominating, a new trend emerges to place the study within the formalism of dynamical systems . In many senses it parallels to the history of advancing from pattern mining of Kepler to mechanistic theory development of Newton in physics and astronomy, though the richness of present-day data is unparalleled. In this review, I only discussed a selection of exampled studies, while the number of such studies grows quickly. For example, Torregrosa and Garcia-Ojalvo reviewed a few more recent studies on analyzing single cell data with the formalism of dynamical systems, which provide mechanistic insights on cell fate decisions in multicellular systems (78). With the early stage of the field, several major challenges remain to be tackled.

First, how to achieve and evaluate completeness of the cell state space description? Mathematically, for a continuous system completeness requires that the dynamical manifold in the observable state space has approximately the same dimensionality of that of the full system, and they are mutually diffeomorphic, so both can be parameterized by the observables (or dynamical variables defined by the observables). Completeness of cell state description is only relative and specific to the cellular process and temporal resolution under consideration. Notice that a cell is a complex system that also interacts with the extracellular environment, therefore any existing or future developed approach can only measure a subset of the degrees of freedom. Fortunately, a dynamical system like a cellular system generally evolves along a low-dimensional dynamical manifold embedded in a high-dimensional state space. The low effective dimensionality of the dynamics makes it theoretically possible for reconstructing the manifolds from incomplete observation of a dynamical system, and various delay and nondelay embedding approaches have been developed (79-84).

Incomplete resolution of cell state leads to distorted description of the cellular dynamics and breakdown of the Markovian assumption. The unexpected detailed balance observed in the study of Chang and Marshall (43) might be due to such incomplete state resolution, especially cell cycle stages. Similarly it is unclear whether the observed two EMT paths of Wang et al (40, 41) truly originate from the same initial fixed-point attractor, or are actually from two groups of distinct but unresolved initial states corresponding to different cell cycle stages, given that one expects a circular instead of a fixed-point attractor if the cell cycle is resolved, and both G1/S and G2/M arrests have been reported during EMT (85-87). To increase the dimension of the state space, in addition to the above embedding methods, some exciting efforts are to increase the modality of the fixed cell data, such as various multi-omics and spatial genomics techniques (27). For live-cell imaging, with advances in multicolor fluorescent imaging and label-free imaging together with deep-learning-based computational approaches (88, 89) one can extract more cellular features from images. Further developments can also include cell-cell interactions and extracellular factors explicitly in the governing equations, and generalize the imaging platform to three-dimensional culturing and in vivo conditions.

Second, how to choose appropriate dynamical variables for reconstructing the governing equations? Choice of coordinate systems is a classical problem, such as using Cartesian versus action-angle coordinates in classical mechanics. With "big data" the problem is also related to dimension reduction. While there are practical constraints on choosing the observables, the dimensionality of the manifold being generally lower than the observable space suggests that a proper choice of the dynamical variables can greatly simplify the form of the governing equations and the number of associated parameters to determine. Most commonly used dimension reduction methods are designed for static data, and dynamical variables obtained through dimension reduction are not necessarily optimal for describing the system dynamics. Lo et al. showed that significance-based variables may not be good predictors (90), and the problem is more prominent for dynamical data. Notice that most dimension reduction methods depend on certain distance measure to quantify, for example, how different the expression states of two cells are. However, a proper distance measure depends on the underlying dynamics. It may take longer time for a cell to transit from one state to another one with "closer" distance than to a different one with a "further" distance under a given measure, if the cellular process needed for the first transition has a slower dynamics than that for the second one. One can understand this situation from the famous fable that when asked by a traveler on the distance to his destination, Aesop replied only after he observed how fast the traveler goes so he could estimate the needed traveling time. Similarly, choosing variables that take into account system dynamics may simplify the equation form and provides more transparent information about the coupling between different components. There are some promising studies on using neural networks to find optimal dynamical variables and reconstruct corresponding governing equations (91, 92). Similarly for image analyses the latent space in a deep neural network can provide variables with interpretable cellular features (93). The Koopman operator, which has been traditionally applied to conservative classical and quantum systems, recently gains increasing popularities in data-driven studies of dynamical systems (94, 95). The Koopman operator, as an extension of the Liouville operator, is the left-adjoint of the Frobenius–Perron operator, and both describe time evolution of densities in the phase space for a deterministic system. Their counterparts for a stochastic system are Kolmogorov's backward and forward equations. With the Koopman operator one can generalize linear analyses on functions into a linear function space composed of nonlinear basis functions (of the observables). Readers can refer to (94) for more systematic discussions.

Third, how to impose physical/biological/dynamical constraints and prior knowledge from other studies into data-driven model reconstruction? Several efforts have been made on approximating cellular dynamics as gradient dynamics. However except in the 1D case (e.g. Fig. 3), one should take caution since a scalar potential does not provide a full description for the dynamics of a driven system. The Helmholtz-Hodge decomposition theorem states that a vector field can be decomposed into a curl-free, a divergent-free, and a harmonic part, and the first two terms can be further defined by a scalar and vector potential, respectively (96). While in many cases a cellular system and a potential system are topologically equivalent in the state space (97, 98), the detailed dynamics can be different (99). The full vector field as reconstructed in Qiu e al. (50) allows systematic investigation on what can and cannot be described with a gradient approximation for a cellular system. With that, the RKHS formalism they used is mathematically convenient, but lacks transparent mechanistic interpretation. It is also challenging to impose physical and biological constraints, such as the requirement that the effect of a gene on another one should reach saturation at high concentrations. This often-conflicting requirement between mathematical convenience and interpretability parallels to the comparison between molecular orbital theory versus valence bond theory in quantum chemistry. It remains to be explored on how to describe gene regulatory network dynamics using function forms familiar to the field of systems biology, and integrate information from scRNA-seq and other types of data, i.e., how to integrate data-driven and mechanism-driven modeling approaches. The efforts can be placed in a larger context of integrating physics- (or biology-) informed machine-learning approaches and mechanistic-oriented research, and Karniadakis et al. discussed several general strategies (100).

Fourth, how to interpret and extract mechanistic information from high-dimensional vector field functions or the governing equations? One excitement is that unbiased description of cellular dynamics provides information on how various cellular processes couple to or insulate from each other, a fundamental problem on understanding the "design principles" of biological networks. A cellular process may be affected by other processes evolving on the same or even slower time scale, leading to nongenetic heterogeneity. This is a phenomenon also discussed in other contexts, named dynamic/static disorders in chemistry, and annealed/quenched disorders in physics (101-103). One expects even richer dynamics with the complexity and nonequilibrium nature of a cellular system. We have seen some concepts and approaches from other established fields introduced to the study of cell phenotypic transitions. For example, the concept of RC has been repetitively adopted in various studies, such as the cell cycle axis in Fig. 1A (49), the popular pseudo-time trajectory analyses in the single cell genomics field (104, 105), and the LAP in Qiu et al. (50). The LAP is the zero-noise (zero-temperature) limit of the RC obtained through the finite temperature string method used in Wang et al. (40), and corresponds to the minimum-energy path for a gradient system. We expect to see applications of other approaches such as the Koopman operator (including dynamic mode decomposition) (94), Zwanzig-Mori projection operator (106-108), in studying cell phenotypic transition dynamics.

In this perspective I discussed analyses with snapshot and live-cell data, each of which has strengths and limitations. The duality and equivalence of representing cell states, in either the expression space (**x**) or cell feature space (**y**), suggest a strategy to combine the strengths of the two (41). The live-cell imaging platform generates real-time single cell trajectories in the cell feature space **y**(t). From the data one can reconstruct the governing equation d**y**/dt = **G**(**y**)+**ξ**(t). Let's assume that the manifolds in the two representations are diffeomorphic or homeomorphic, provided that we have complete description of cell states in both representations, while this one-to-one mapping is convenient but not necessary. Then by establishing a mapping between the cell state in the cell feature space and expression space, M:**y**→**x**, one can obtain the kinematics description and the governing equation d**x**/dt = **F**(**x**)+**η**(t) in the expression space that contains mechanistic information on gene regulation. The feasibility of the proposal may seem questionable given the seemingly mismatch on the dimensionality of fixed cell data one can acquire and that of the live-cell imaging data. The underlying theoretical consideration is that a high-dimensional dynamical system typically evolves along a low-dimensional manifold due to the constraints imposed by the component interactions. It is exciting to see whether this integrated strategy can lead to a more complete description of cellular dynamics, specifically cell phenotypic transitions.

## Acknowledgments

I thank Drs Ivet Bahar, Allon Klein, Sachit Dinesh Saksena, David Gifford, Xiaojie Qiu, as well as the anonymous reviewers for reading the manuscript and providing constructive comments. This work was partially supported by National Cancer Institute (R37 CA232209), National Institute of Diabetes and Digestive and Kidney Diseases (R01DK119232), and National Science Foundation (2205148) to JX.